\documentclass[twocolumn]{aastex7}
\usepackage{amsmath,amsfonts,amssymb, mathtools}

\newcommand{\vect}[1]{\boldsymbol{\mathbf{#1}}}

\shortauthors{Chen et al.}
\graphicspath{{./}{./figures/}}

\begin{document}

\title{Stacking 21-cm Maps around Lyman-$\alpha$ Emitters during Reionization: \\Prospects for a Cross-correlation Detection with the Hydrogen Epoch of Reionization Array}

\author[0000-0002-3839-0230]{Kai-Feng Chen}
\affiliation{MIT Kavli Institute for Astrophysics and Space Research, Cambridge, MA 02139, USA}
\affiliation{Department of Physics, Massachusetts Institute of Technology, Cambridge, MA 02139, USA}
\email[show]{kfchen@mit.edu}

\author[0000-0002-9205-9717]{Meredith Neyer}
\affiliation{MIT Kavli Institute for Astrophysics and Space Research, Cambridge, MA 02139, USA}
\affiliation{Department of Physics, Massachusetts Institute of Technology, Cambridge, MA 02139, USA}
\email{mneyer@mit.edu}

\author[0000-0002-4117-570X]{Jacqueline N. Hewitt}
\affiliation{MIT Kavli Institute for Astrophysics and Space Research, Cambridge, MA 02139, USA}
\affiliation{Department of Physics, Massachusetts Institute of Technology, Cambridge, MA 02139, USA}
\email{jhewitt@mit.edu}

\author[0000-0002-2838-9033]{Aaron Smith}
\affiliation{Department of Physics, The University of Texas at Dallas, Richardson, TX 75080, USA}
\email{asmith@utdallas.edu}

\author[0000-0001-8593-7692]{Mark Vogelsberger}
\affiliation{MIT Kavli Institute for Astrophysics and Space Research, Cambridge, MA 02139, USA}
\affiliation{Department of Physics, Massachusetts Institute of Technology, Cambridge, MA 02139, USA}
\affiliation{The NSF AI Institute for Artificial Intelligence and Fundamental Interaction, Cambridge, MA 02139, USA}
\email{mvogelsb@mit.edu}


\begin{abstract}
Observations of the redshifted 21-cm line during the Epoch of Reionization will open a new window to probe the intergalactic medium during the formation of the first stars, galaxies, and black holes. A particularly promising route to an initial detection is to cross-correlate tomographic 21-cm maps with spectroscopically confirmed Lyman-$\alpha$ emitters (LAEs). High-redshift LAEs preferentially reside in ionized bubbles that are strongly anticorrelated with the surrounding neutral regions traced by 21-cm observations. In this work, we study the prospect of detecting such a cross-correlation signal by stacking 21-cm image cubes around LAEs using a current-generation 21-cm instrument---the Hydrogen Epoch of Reionization Array (HERA). Our forecast adopts a realistic mapping pipeline to generate foreground-free 21-cm image cubes. The statistical properties of these images, arising from the complex instrumental response, are carefully accounted for. We further introduce a physically motivated signal template calibrated on the \textsc{thesan} radiation-hydrodynamic simulations, which connects the cross-correlation amplitude to the global neutral fraction. Our results show that a sample of $\sim$50 spectroscopically confirmed LAEs is sufficient to begin constraining the reionization history. These results represent an important preparatory step toward joint analyses of 21-cm experiments with upcoming wide-area, high-redshift galaxy surveys from \textit{Euclid} and the \textit{Nancy Grace Roman Space Telescope}.
\end{abstract}



\section{Introduction} 

The Epoch of Reionization (EoR) is an astrophysically complex era that has yet to be fully explored \citep{Loeb_Barkana2001:ARANA, Robertson2022:ARANA}. Although the midpoint of reionization has been constrained by cosmic microwave background experiments \citep[e.g.,][]{Planck18, Pagano2020:Planck_optical_depth, Li2025:CLASS}, and the reionization history has also been constrained from various quasar sightline observations \citep[e.g.,][]{Durovcikova2020:Damping_wing_reionization, Wang2021:highz_quasar, Becker2021:QSO_reion, Bosman2021:Comparison, Bosman2022:QSO_reion, Gaikwad2023:QSO_reionization, Durovcikova2024:Damping_wing_reionization}, recent observations of high-redshift galaxies---especially the discovery of numerous Lyman-$\alpha$ emitters (LAEs) deep in reionization---have raised questions about the detailed processes of reionization \citep{Umeda2024:LAEs_Damping_Wings, Finkelstein2024:CEERS, Munoz2024}. A deeper understanding of the EoR will not only provide rich astrophysical insights into the properties of the first galaxies but also yield significant cosmological implications \citep{McQuinn2006:21cm_Neutrino, Liu2016:21cm_Neutrino, Sailer2025:low_tau, Jhaveri2025:low_tau_neutrino, Allali2025:tau_H0}. 

The redshifted 21-cm line from neutral hydrogen provides a direct and comprehensive probe of this otherwise opaque period in cosmic history \citep{Furlanetto2006:Review, Pritchard2012:Review}. The current generation of large radio interferometers, such as the Hydrogen Epoch of Reionization Array (HERA, \citealt{HERA2017:PhaseI_Overview, Berkhout2024:HERA_PhaseII}), has already been setting stringent limits on the 21-cm power spectrum---a statistical measurement of the spatial fluctuations in the 21-cm signal---and has placed important constraints on the properties of the intergalactic medium (IGM) during reionization \citep{HERA2022:h1c_idr2_limit, HERA2023:h1c_idr3_limit}. Continued observations with current experiments such as HERA and the upcoming Square Kilometre Array (SKA, \citealt{SKA2015:EoR}) will soon reach sufficient sensitivity to detect the 21-cm auto-spectrum \citep{Breitman2024:HERA_Forecast}.

Meanwhile, large ensembles of high-redshift galaxies discovered by ground- and space-based instruments \citep{Malhotra2004:LAEs_LF, Ouchi2008:Subaru_LF, Ouchi2010:Subaru_LF, Konno2018:Subaru_LF, Zheng2017:LAGER_Overview, Wold2022:LAE_LF, Kumari2024:JADES_LAE} enable an alternative route to detecting the 21-cm signal via cross-correlation \citep{LaPlante2023:21cm_x_Roman, Gagnon-Hartman2025:21cm_x_LAE, Hutter2025:Cross_Correlation}. Under an ``inside-out'' reionization scenario, the 21-cm signal anticorrelates with the galaxy field: overdense regions around galaxies reionize first, while the 21-cm signal continues to trace the surrounding neutral regions \citep{Choudhury2009, Kannan2022:THESAN_LIM}. Thanks to the high signal-to-noise nature of optical observations, such an anticorrelation could be easier to detect. This is evident from the fact that most cosmological measurements of 21-cm fluctuations in the post-reionization universe have been made through correlating with galaxy surveys \citep{Chang2010:Cross_correlation, Masui2013:Cross_correlation, CHIME2023:Cross_correlation, CHIME2024:Cross_correlation}. If detected, this cross-correlation will serve as a crucial sanity check for any future 21-cm auto-spectrum detection.

Here, we study the prospect of detecting a cross-correlation signal by stacking 21-cm image cubes around LAEs. Although many studies have investigated cross-correlating 21-cm data with galaxies \citep{Hutter2017:21cm_LAE, Heneka2021:21cm_LAE, Davies2021:21cm_stack, Cox2022:cross-correlation, LaPlante2023:21cm_x_Roman, Gagnon-Hartman2025:21cm_x_LAE, Hutter2025:Cross_Correlation}, most focused on correlation functions or cross power spectra. A stacking signal contains less information as full power spectra, detecting such a signal during the EoR still has significant astrophysical implications, because the signal strength is directly sensitive to the global $\mathrm{HI}$ fraction \citep{Hutter2023:21cm_LAE}. 

In this work, we improve on previous studies by accounting for additional observational and theoretical complexities. We utilize a direct optimal mapping framework \citep{Xu2022:DOM}, which allows us to accurately quantify the statistical properties of these 21-cm image cubes. A realistic foreground-filtering algorithm \citep{Ewall-Wice2021:DAYNENU} is applied to quantify signal loss from foreground mitigation. On the theory side, we use the radiation-magneto-hydrodynamics simulations \textsc{thesan} \citep{Garaldi2022:THESAN_IGM, Kannan2022:THESAN_Overview, Kannan2025:THESAN_zoom, Smith20222:THESAN_Lya} to derive a signal template. The full radiative transfer calculations adopted within \textsc{thesan} provide a robust physical connection between galaxies and ionized bubbles \citep{Kannan2022:THESAN_LIM, Yeh2023:THESAN_fesc, Neyer2024:THESAN_bubble, Jamieson2025, Zhao2025}, including analyses specifically targeting LAE populations \citep{Smith20222:THESAN_Lya, Xu2023:Thesan_Lya}. This allows us to account for the complex correlation between the optical properties of LAEs and their surrounding IGM as observed in radio. 

This paper is organized as follows. In Sec.\,\ref{sec:mapping} we discuss the procedure to generate foreground-filtered 21-cm image cubes and their statistical properties. A theory template for the cross-correlation signal inferred from simulations is given in Sec.\,\ref{sec:theory_template}. The prospects for a cross-correlation detection and its cosmological implications are discussed in Sec.\,\ref{sec:results}. Conclusions are given in Sec.\,\ref{sec:conclusion}.

\section{Foreground filtered 21-cm Maps}\label{sec:mapping}

In this work, we adopt the direct optimal mapping\footnote{\url{https://github.com/HERA-Team/direct_optimal_mapping}} framework developed in \citet{Xu2022:DOM} to generate 21-cm image cubes. This formalism is particularly beneficial for our application, as the statistical properties of the images are well understood. Here, we provide a brief overview of direct optimal mapping in Sec.\,\ref{subsec:DOM}. We describe how we filter foreground contamination from the data in Sec.\,\ref{subsec:foreground_filter}. The statistical properties of these maps are presented in Sec.\,\ref{subsec:statistics}.

\subsection{Direct Optimal Mapping}\label{subsec:DOM}
The most natural data product from a radio interferometer is the correlation of voltages measured by any two antennas, \textit{i.e.}, the \textit{visibility}, as a function of frequency $\nu$, 
\begin{equation}
\label{eq:measurement_eq}
    V(\mathbf{b}_{ij}, \nu) = \int \mathrm{d}\Omega\,I(\hat{\mathbf{s}}, \nu) B_{ij}(\hat{\mathbf{s}}, \nu) \exp\left(-\frac{i2\pi\nu}{c}\mathbf{b}_{ij}\cdot\hat{\mathbf{s}}\right)\, .
\end{equation}
Here, $i$ and $j$ are antenna indices; $I$ is the brightness temperature of the sky; $B_{ij}$ is the cross-power beam; $\mathbf{b}_{ij}$ is the baseline vector; and $\hat{\mathbf{s}}$ is the unit vector on the sky over which we integrate. 

We can discretize Eq.\,\eqref{eq:measurement_eq} and describe the relation between the interferometric data $\vect{d}$ and the sky $\vect{m}$ using a linear system. For a given time and frequency, we write,
\begin{equation}
    \vect{d} = \vect{A}\vect{m} + \vect{n}\, .
\end{equation}
Here, $\vect{d}$ is a vector with a dimension equal to the number of baselines, and $\vect{m}$ is a vector with a dimension equal to the number of discretized sky pixels. $\vect{n}$ represents instrumental noise and can also absorb other uncertainties such as discretization error\footnote{In this work, we choose a pixelization scheme that has much higher resolution than our assumed instrument. We therefore assume any discretization error is negligible.}. The design matrix $A$ is written as 
\begin{equation}
\label{eq:amat}
    A_{mn} \equiv B_{m}(\hat{\mathbf{s}}_n)\exp\left(-\frac{i2\pi\nu}{c}\mathbf{b}_{m}\cdot\hat{\mathbf{s}}_n\right)\, ,
\end{equation}
where the index $m$ runs over the baseline axis and $n$ runs over the sky pixel axis. We note that we absorb the area element $\Delta\Omega$ from Eq.\,\eqref{eq:measurement_eq} into the sky vector $\vect{m}$. While $I(\hat{\mathbf{s}}, \nu)$ has a unit of specific intensity (e.g., $\left[\mathrm{Jy}/\mathrm{Sr}\right]$), we choose to work with $\vect{m}$ representing the flux density from each pixel (e.g., in units of $\left[\mathrm{Jy}\right]$). 

A simple but sufficient estimator for the true sky $\vect{m}$ can be formed as
\begin{equation}
\label{eq:mapping_from_vis}
    \vect{\hat{m}} \equiv \vect{D}\vect{A}^\dagger \vect{N}^{-1}\vect{d}\, ,
\end{equation}
where $\vect{D}$ is a normalization matrix and $\vect{N} \equiv \langle n n^\dagger\rangle$ is the noise covariance. $\vect{\hat{m}}$ satisfies the Fisher–Neyman criterion as long as $\vect{D}$ is invertible \citep{Tegmark1997:Optimal_mapping}. Throughout this work, we assume that the noise covariance in visibility space follows the form 
\begin{equation}
\label{eq:noise_matrix}
    N_{ij} = \frac{\sigma^2_\mathrm{rms}(\nu, t)}{n_{\vect{b}_i}} \delta_{ij}\, ,
\end{equation}
where $\sigma_\mathrm{rms}$ is given by the radiometer equation \citep{Tan2021:HERA_ErrorBar},
\begin{equation}
\label{eq:radiometer}
    \sigma_\mathrm{rms} \equiv \frac{T_\mathrm{sys}}{\sqrt{\Delta\nu\Delta t}}\, .
\end{equation}
Here, $n_{\vect{b}_i}$ is the number of redundancy for each baseline group, $T_\mathrm{sys}$ is the system temperature which is often estimated from the antenna's auto-correlation, $\Delta \nu$ is the correlator channel width, and $\Delta t$ is the correlator integration time. For HERA, $\Delta \nu = 122\,\mathrm{kHz}$ and $\Delta t = 9.6\,\mathrm{s}$.

\subsection{Imaging Delay-filtered Visibilities} \label{subsec:foreground_filter}

One of the major barriers for 21-cm cosmology is the presence of bright foreground emission combined with complex instrumental responses. For a radio interferometer, foreground contamination is usually confined to a region of Fourier space known as the foreground wedge \citep{Datta2010:FGwedge, Parsons2012:delay_spectrum_wedge, Vedantham2012:image_wedge, Trott2012:wedge, Morales2012:wedge, Hazelton2013:wedge, Thyagarajan2013:wedge, Liu2014:EoR_WindowI}. While there exists a rich literature in explicitly modeling and subtracting the foregrounds, foreground subtraction remains challenging in practice, especially in the presence of uncertainties in instrument response and systematic effects. Therefore, the most conservative method of mitigating foreground is to filter out all modes within the foreground wedge.

Ideally, for the visibility measured by each baseline $\vect{b}$, the foreground contamination should predominantly reside within delay $|\tau| \leq |\vect{b}|/c$, where $c$ is the speed of light and $\tau$ is the Fourier dual of the frequency for each baseline. Here, we adopt a foreground filtering method first developed in \cite{Ewall-Wice2021:DAYNENU}, utilizing a set of basis functions known as the discrete prolate spheroidal sequence (DPSS, \citealt{Slepian1978:DPSS}). The smooth foreground component in each visibility is removed by fitting these basis functions that are localized in Fourier space (within $|\tau| \leq |\vect{b}|/c$). For details of this procedure, we refer the reader to \autoref{appendix:fg_filter}. For the purpose of this work, we simply treat foreground filtering as a linear operation $\vect{\mathcal{O}^\mathrm{fil}}$, in which the filtered visibility is obtained via
\begin{equation}
\label{eq:fg_filt}
    V^\mathrm{fil}(\mathbf{b}, \nu_i) = \sum_j\mathcal{O}^\mathrm{fil}_{ij}(\mathbf{b}) V(\mathbf{b}, \nu_j)\, .
\end{equation}
This foreground-filtering method has proven successful in real-world applications \citep{CHIME2023:21cm_x_QSO_ELG, CHIME2023:21cm_x_Lya, HERA2023:h1c_idr3_limit, HERA2025:h6c_idr2_limit} and is particularly beneficial in dealing with data with gaps, often introduced by radio frequency interference \citep{Chen2025:Inpainting}. While residual foreground may persist due to systematic effects \citep[e.g.,][]{Kim2022:BeamPerturbation, Pascua2024:FRF, Rath_Pascua2024:Mutual_Coupling}, we defer these to future studies. For the remainder of this work, we assume foreground contamination is completely removed by this delay-filtering procedure.

\subsection{Noise Properties}\label{subsec:statistics}

Because the mapping and foreground filtering procedures outlined in Sec.\,\ref{subsec:DOM} and \ref{subsec:foreground_filter} are both linear, a key advantage of this framework is that the noise properties of the image cubes can be easily modeled. Here, we discuss a couple of important statistical properties of our image cubes, including the choice of normalization, the optimal time averaging procedure, and both frequency-frequency and pixel-pixel correlations. We note that while the mathematical framework presented here is generic, the plots shown in this section require specifying an array layout. For all results this section, we use the portion of the HERA array commissioned as of 2022 as an example. We refer the reader to Sec.\,\ref{subsec:forecast_setup} for more details. 

First, we discuss the choice of a sensible normalization convention. We note that the (pixel-pixel) noise covariance for each map is 
\begin{equation}
\label{eq:noise_img}
\begin{aligned}
    \vect{N}^\mathrm{img} &\equiv \langle \vect{\hat{m}} \vect{\hat{m}}^\dagger\rangle \\
    &= \vect{D}\vect{A}^\dagger \vect{N}^{-1} \langle \vect{n} \vect{n}^\dagger\rangle \vect{N}^{-1} \vect{A} \vect{D}^\dagger \\
    &= \frac{1}{2} \vect{D}\vect{A}^\dagger \vect{N}^{-1} \vect{A} \vect{D}^\dagger\, ,
\end{aligned} 
\end{equation}
where the factor of $1/2$ arises because we only take the real part of the image. Without the normalization matrix $\vect{D}$, the noise variance in each sky pixel $\vect{\hat{s}}_i$ is proportional to

\begin{equation}
\label{eq:map_pixel_variance}
\begin{aligned}
    \left(\vect{A}^\dagger \vect{N}^{-1} \vect{A}\right)_{ii} &= |B(\vect{\hat{s}}_i)|^2 \sum_k  N^{-1}_{kk} \\
    &=\frac{n_\mathrm{bl}|B(\vect{\hat{s}}_i)|^2}{\sigma_\mathrm{rms}^2}\, , 
\end{aligned}
\end{equation}
where we have assumed cross power beams are the same across all baselines. Here, $n_\mathrm{bl}$ is the total number of baselines in the array and $\sigma_\mathrm{rms}$ is given by Eq.\,\eqref{eq:radiometer}. Therefore, a natural choice for $\vect{D}$ so that the noise property is uniform across all pixels is to have $D_{ij} \propto \delta_{ij}/B(\vect{\hat{s}}_i)$. 

On the signal side, our image estimator $\vect{\hat{m}}$ relates to the true sky $\vect{m}$ via
\begin{equation}
        \langle\vect{\hat{m}}\rangle = \vect{D}\vect{A}^\dagger \vect{N}^{-1}\vect{A}\vect{m}\, . 
\end{equation}
In particular, in this work we focus on the response of the image estimator to a point source in the sky. While reionization bubbles usually span several megaparsecs, current radio instruments such as HERA do not have the sensitivity to spatially resolve them. Therefore, we treat these bubbles as point sources and focus on obtaining the cross-correlation signal along the frequency direction, where radio interferometers have exquisite resolution. For a point source centered on a sky pixel $\vect{\hat{s}}_i$, we have
\begin{equation}
\label{eq:pt_source_signal}
    \langle{\hat{m}_i}\rangle = D_{ii}\frac{n_\mathrm{bl}|B(\vect{\hat{s}}_i)|^2}{\sigma_\mathrm{rms}^2} f_i\, ,
\end{equation}
where $f_i$ is the flux of the point source and we have assumed the normalization matrix $\vect{D}$ is diagonal. Hence, another choice of $\vect{D}$ so that the peak flux of any point source is preserved is to have $D_{ij} = \delta_{ij} \sigma^2_\mathrm{rms} / n_\mathrm{bl} / |B(\vect{\hat{s}}_i)|^2$. However, as shown in Eq.\,\eqref{eq:map_pixel_variance}, this normalization increases the noise variance for pixels farther from the pointing center.

In this work, we choose the following normalization matrix, 
\begin{equation}
\label{eq:normalization}
    D_{ij} = \frac{\sigma_\mathrm{rms}^2}{n_\mathrm{bl} |B(\vect{\hat{s}}_i)|}\delta_{ij}\, .
\end{equation}
Plugging this into Eq.\,\eqref{eq:pt_source_signal}, such a normalization gives rise to a beam-weighted sky. As discussed above, this choice of normalization also yields a uniform noise variance across all pixels. A clear advantage of this approach is that the mapping between visibilities to maps according to Eq.\,\eqref{eq:mapping_from_vis} does not depend on our knowledge of the primary beam, since the beam factor cancels between Eq.\,\eqref{eq:amat} and Eq.\,\eqref{eq:normalization}. In reality, beam modeling can be complex and uncertain. This approach ensures that uncertainties in beam modeling do not propagate into calculations of noise statistics but are instead contained entirely within the signal modeling. 

Because the current generation of radio interferometers lacks the sensitivity to detect individual ionized bubbles, it is crucial to correctly average the data along different axes to increase sensitivity. This can be done by averaging image cubes from different times or by stacking different pixels that contain ionized bubbles. However, extra care is required in the presence of pixel-pixel correlations and time-dependent noise properties. 

\begin{figure}[tbh]
    \centering
    \epsscale{1.175}
    \plotone{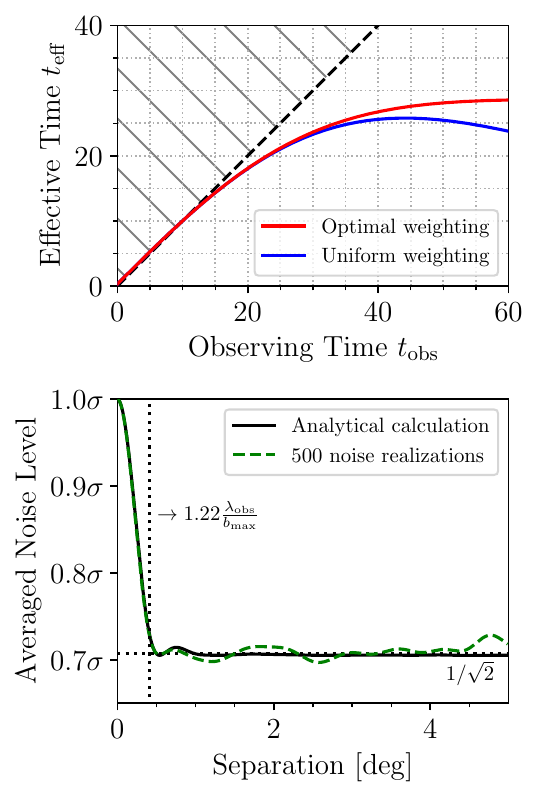}
    \caption{$\textit{Top}$: Effective observing time (see Eq.\,\ref{eq:t_eff}) for a transit array under different time-average weighting scheme. This shows how the signal-to-noise ratio for a point source changes (measured in terms of $\sqrt{t_\mathrm{eff}}$) as a function of observed time $t_\mathrm{obs}$. Here, we assume the source transits across zenith at $t_\mathrm{obs}=0$. \textit{Bottom}: Averaged noise level when stacking two lines of sight from different separation. If two lines of sight are completely independent, the noise level should decrease by a factor of $\sqrt{2}$. Instrumental response makes pixels that lie within the array's \textit{synthesis beam} particularly correlated with each other. This is well characterized by the diffraction limit scale $1.22\,\lambda_\mathrm{obs} / b_\mathrm{max}$. In this plot, the observed wavelength $\lambda_\mathrm{obs}\approx 1.67\,\mathrm{meter}$ which traces 21-cm lines at $z\sim7$, and the longest baseline length is at $b_\mathrm{max}\approx265\,\mathrm{meter}$. \label{fig:avg_statistics}} 

\end{figure}

First of all, we consider coherent time averaging by mapping the same pixel with visibilities obtained at different times. For a tracking telescope, the noise variance decreases as $1/t_\mathrm{obs}$. However, for a transit telescope like HERA, as the source moves across the antenna's primary beam, the accumulated signal-to-noise ratio is different depending on the location of the source. In other words, a visibility measured at a local sidereal time far from the source’s right ascension contributes little signal to the image. Combining Eq.\,\eqref{eq:pt_source_signal} and \eqref{eq:map_pixel_variance}, the signal-to-noise ratio of a point source at $\vect{\hat{s}}_i$ with flux $f_i$ observed at time $t$ is 
\begin{equation}
\label{eq:pt_source_SN}
    \mathcal{S}/\mathcal{N}(t) = \frac{\sqrt{2n_\mathrm{bl}}}{\sigma_\mathrm{rms}}|B(\vect{\hat{s}}_i; t)|f_i \, , 
\end{equation}
which is independent from the choice of the normalization matrix. Because of this, although we adopt a normalization that ensures uniform noise levels over time, simply averaging visibilities uniformly across different times does not maximize the signal-to-noise ratio. The signal is diluted as the source moves away from zenith. Here, we consider two averaging schemes $w_i(t)$: a uniform weighting scheme where $w_i(t) \propto 1$ and an \textit{optimal} weighting scheme in which $w_i(t) \propto |B(\vect{\hat{s}}_i; t)|$. To assess how time-averaging helps increase the signal-to-noise ratio, we define an effective integration time $t_\mathrm{eff}$, given a period of observation $t_\mathrm{obs}$, to be
\begin{equation}
\label{eq:t_eff}
\begin{aligned}
        &t_\mathrm{eff}(t_\mathrm{obs}) =\\
        &\left(\frac{1}{\mathcal{S}/\mathcal{N}(t_0)}\frac{\int_{t_0 - t_\mathrm{obs}/2}^{t_0 + t_\mathrm{obs}/2} w_i(t) \langle{\hat{m}_i}\rangle(t)~\mathrm{d}t}{\sqrt{\int_{t_0 - t_\mathrm{obs}/2}^{t_0 + t_\mathrm{obs}/2} w_i(t)N^\mathrm{img}_{ii}(t)~\mathrm{d}t}}\right)^2\, ,
\end{aligned}
\end{equation}
where $t_0$ is the time where the source transits the zenith. For a tracking telescope with uniform weighting, $t_\mathrm{eff}(t_\mathrm{obs}) = t_\mathrm{obs}$. The top panel of \autoref{fig:avg_statistics} shows $t_\mathrm{eff}$ for the two weighting schemes considered above for a HERA-like transit experiment. We see that the signal-to-noise ratio under the \textit{optimal} weighting scheme (red) saturates after a given time and reaches a higher signal-to-noise ratio compared to the naive uniform weighting scheme (blue). The signal-to-noise ratio for the uniform weighting scheme drops after a certain time as the averaging process contributes more noise than signal. Although an \textit{optimal} weighting scheme achieves around $10\%$ higher signal-to-noise ratio from averaging, it requires precise knowledge of the instrument's primary beam. In this work, we adopt the more conservative and practical uniform-averaging scheme for our forecast. Therefore, for each night of observations, we assume that we can coherently average the image for around 45 minutes to reach an effective average time of roughly 25 minutes. The latter quantity is often referred to as the \textit{beam crossing time}. We note that for simplicity, here we have assumed that the instrument is stable as a function of time, \textit{i.e.}, the number of available baselines, the system temperature, and the antenna's primary beam are all not a function of time. 

Another way to increase the signal-to-noise ratio for cross-correlation is by stacking 21-cm image cubes around known LAEs. If each line of sight is completely independent, then stacking around $N$ galaxies reduces the noise by $\sqrt{N}$. However, as seen from Eq.\,\eqref{eq:noise_img}, there exists non-trivial pixel-pixel correlations due to instrumental response. The noise variance for each pixel is the same,
\begin{equation}
\label{eq:pixel_variance}
    \sigma^2=\frac{\sigma^2_\mathrm{rms}}{2\,n_\mathrm{bl}}\, ,
\end{equation} 
thanks to our choice of the normalization matrix. The correlation between two pixels $i$ and $j$ is given by
\begin{equation}
\label{eq:pixel_corr}
    \rho_{ij} \equiv \frac{N^\mathrm{img}_{ij}}{\sqrt{N^\mathrm{img}_{ii}N^\mathrm{img}_{jj}}}\, .
\end{equation}
The resulting noise variance after averaging two lines of sight is 
\begin{equation}
\label{eq:avg_variance}
    \frac{1+\rho_{ij}}{2}\sigma^2\, .
\end{equation}
The bottom panel of \autoref{fig:avg_statistics} shows the agreement between an analytical calculation and numerical noise simulations. The solid black line is calculated using Eq.\,\eqref{eq:avg_variance}, while the dashed red line is obtained from $500$ realizations of noise simulations. For each noise realization, we generate images at different pointings (constant declination, separated by right ascension) from noise-only visibilities according to Eq.\,\eqref{eq:radiometer}. Each pointing is then time-averaged for $45$ minutes to reach maximum sensitivity. We see that pixels within the array's \textit{synthesis beam} are particularly correlated with each other. This is well characterized by the diffraction limit scale $1.22\,\lambda_\mathrm{obs} / b_\mathrm{max}$. In this plot, the observed wavelength $\lambda_\mathrm{obs}\approx 1.67\,\mathrm{meter}$ traces 21-cm lines at $z\sim7$, and the longest baseline length is at $b_\mathrm{max}\approx265\,\mathrm{meter}$. Therefore, if we stack $N$ galaxies that are pairwise separated by at least the diffraction limit, we can safely assume that the noise level decreases by $\sqrt{N}$. 

\begin{figure}[tbh]
    \centering
    \epsscale{1.175}
    \plotone{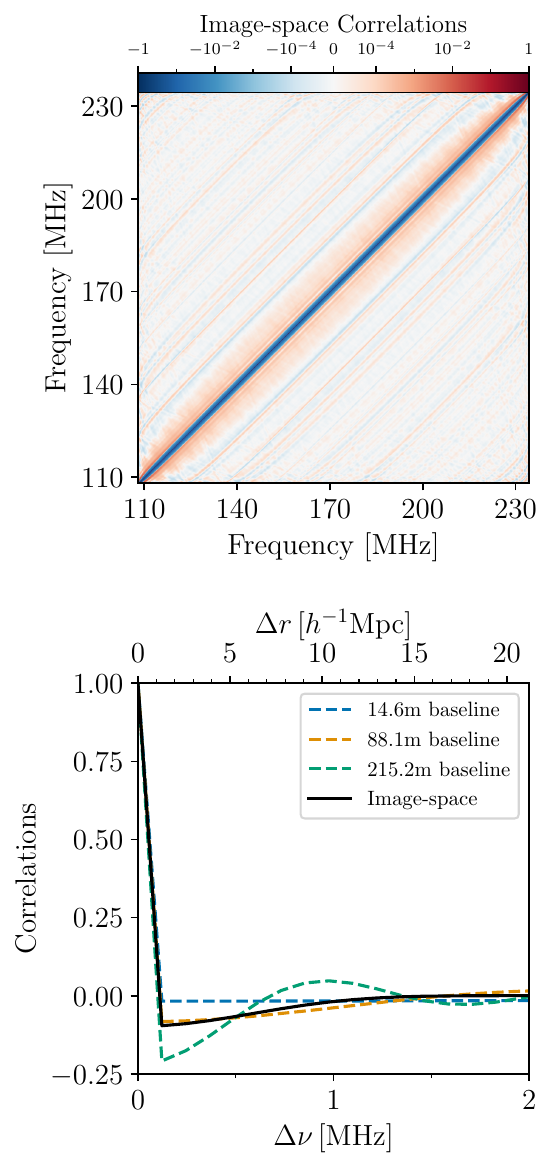}
    \caption{$\textit{Top}$: Frequency-frequency correlations in image space due to foreground filtering. Here, each baseline is filtered to the delay of $|\vect{b}|/c$. \textit{Bottom}: Correlation in image space (solid black line) versus those in visibility space for various baselines (dashed lines). Here, we are showing correlation between the frequency channel at around $179.9$ MHz (which traces 21-cm lines at $z\sim7$) with its neighboring channels. \label{fig:freq_freq_statistics}} 

\end{figure}

So far, we have focused on the image properties at one frequency. While images at different frequency bins should be independent, the foreground filtering procedure described in Sec.\,\ref{subsec:foreground_filter} introduces frequency-frequency correlations in each single-baseline visibility. Here, we investigate how optimal mapping propagates visibility-space correlations into image-space correlations. 

Following Eq.\,\eqref{eq:fg_filt}, after foreground filtering, the visibility measured by a given baseline is correlated between different frequency channels through
\begin{equation}
\begin{aligned}
    \left(\vect{C}^\mathrm{filt}_{\vect{b}}\right)_{ij} &\equiv \langle V^\mathrm{fil}(\mathbf{b}, \nu_i) V^\mathrm{fil}(\mathbf{b}, \nu_j)^*\rangle \\ 
    &= \sum_{mn}\mathcal{O}^\mathrm{fil}_{im}(\mathbf{b}) V(\mathbf{b}, \nu_m)\mathcal{O}^\mathrm{fil}_{jn}(\mathbf{b})^* V(\mathbf{b}, \nu_n)^*    \\
    &= \left(\vect{\mathcal{O}}^\mathrm{fil}(\mathbf{b})\vect{C}_{\vect{b}}\vect{\mathcal{O}}^\mathrm{fil}(\mathbf{b})^\dagger\right)_{ij} \, ,
\end{aligned}
\end{equation}
where we assume that the data covariance $\vect{C}_{\vect{b}}$ is noise dominant,
\begin{equation}
    \left(\vect{C}_{\vect{b}}\right)_{ij}\equiv \frac{\sigma^2_\mathrm{rms}(\nu_i)}{n_{\vect{b}}} \delta_{ij}\, .
\end{equation}
As our foreground-filtering method does not correlate between different baselines, the corresponding frequency-frequency covariance for a given line of sight $\vect{\hat{\mathbf{s}}}_n$ is 
\begin{equation}
\label{eq:img_covariance}
    \begin{aligned}
        &\left(\vect{C}^\mathrm{img}_{\vect{\hat{\mathbf{s}}}_n}\right)_{ij} \equiv \langle \hat{m}_n(\nu_i) \hat{m}_n(\nu_j)^*\rangle \\
        =& \sum_k \left(\frac{n_{\vect{b}_k}}{n_\mathrm{bl}}\right)^2 \exp\left(\frac{i2\pi(\nu_i-\nu_j)}{c}\vect{b}_k\cdot\vect{\hat{\mathbf{s}}}_n\right)\left(\vect{C}^\mathrm{filt}_{\vect{b}}\right)_{ij}\, .
    \end{aligned} 
\end{equation}
The image-space frequency-frequency covariance is simply a weighted sum of the visibility-space covariance across all baselines. 

The top panel of \autoref{fig:freq_freq_statistics} shows the frequency-frequency correlations, $\vect{C}^\mathrm{img}_{\vect{\hat{\mathbf{s}}}_n}/\sigma^2$, across the entire frequency band where we use to filter the foreground. Although ionized bubbles are fairly localized in frequency space, we still choose a wide frequency band because this minimizes signal attenuation during foreground filtering \citep{Ewall-Wice2021:DAYNENU, Kern2021:GPR}. For longer baselines, this introduces long-range correlations as more line-of-sight modes are filtered (e.g., the dashed green line in the bottom panel of \autoref{fig:freq_freq_statistics}). However, since the image-space correlation is a linear combination of visibility-space correlations across all baselines weighted by the redundancy of each baseline group, it is dominant by the behavior of shorter baselines. The bottom panel of \autoref{fig:freq_freq_statistics} shows that there is in fact no significant correlation in the image space beyond its immediate neighboring frequency channels. 

\section{Signal Modeling}\label{sec:theory_template}
To accurately forecast and interpret the cross-correlation signal, we use the radiation-magneto-hydrodynamic simulations \textsc{thesan} to derive a signal template which takes into account the selection effects of LAEs. In the following, we give a brief overview of the \textsc{thesan} simulation in Sec.\,\ref{subsec:THESAN}. Observed properties of LAEs are discussed in Sec.\,\ref{subsec:LAE_properties}. Combining these, we present the modeling template for the cross-correlation signal in Sec.\,\ref{subsec:signal_template}. 

\subsection{Simulations} \label{subsec:THESAN}

\textsc{Thesan} \citep{Kannan2022:THESAN_Overview, Smith20222:THESAN_Lya, Garaldi2022:THESAN_IGM} is a suite of large (95.5 comoving Mpc per side) radiation-magneto-hydrodynamic cosmological simulations run down to $z = 5.5$, which model reionization by self-consistently combining on-the-fly radiative transfer and realistic galaxy formation modeling from IllustrisTNG \citep{Vogelsberger2013:Illustris, Vogelsberger2014a:Illustris, Vogelsberger2014b:Illustris, Weinberger2017:IllustrisTNG, Pillepich2018a:IllustrisTNG, Springel2018:IllustrisTNG, Vogelsberger2020:Review}. Here, we adopt the fiducial simulation \textsc{thesan-1}, which resolves dark matter to $3.1 \times 10^6\, \text{M}_\odot$ and baryonic matter to $5.8 \times 10^5\, \text{M}_\odot$. Atomic cooling halos are therefore marginally resolved down to masses of $M_{\rm halo} \gtrsim 10^8\, h^{-1}\, \text{M}_\odot$. \textsc{Thesan} uses the efficient quasi-Lagrangian code \textsc{arepo-rt} \citep{Kannan2019:AREPO-RT, Zier2024:AREPO}, an extension of the moving mesh code \textsc{arepo} \citep{Springel2010:AREPO, Weinberger2020:AREPO}, with additional physics required to self-consistently model reionization. It solves the fluid dynamics equations on an adaptive unstructured Voronoi mesh produced by approximately following the flow of the gas. Gravity calculations utilize a hybrid Tree-PM approach, which splits the force into short- and long-range contributions \citep{Barnes1986:TreePM}. The radiation transport equations are solved using a moment-based approach assuming the M1 closure relation \citep{Levermore1984, Dubroca1999}, with the spectrum discretized in three frequency ranges to accurately capture non-equilibrium photoionization and photoheating from stellar and AGN sources for primordial gas. A reduced speed of light approximation is used with an effective value of $0.2\,c$, and a birth cloud escape fraction of $0.37$ is employed to match constraints for the global reionization history. Data products from the \textsc{thesan} simulations are publicly available online for community use \citep{Garaldi2023:THESAN_Data_Release}.

To obtain the properties of the 21-cm field from \textsc{thesan}, we model its brightness temperature via \citep{Furlanetto2006:Review}
\begin{equation}
\label{eq:21cm_brightness}
\begin{aligned}
        \delta T_b \approx& 27 \,{\rm mK} (1+\delta_b)x_{\rm HI}\left(1-\frac{T_{\rm CMB}(\nu)}{T_{\rm spin}}\right)\left(\frac{\Omega_b h^2}{0.023}\right) \\ 
        &\times\sqrt{\left(\frac{1+z}{10}\right)\left(\frac{0.15}{\Omega_m h^2}\right)}\left(\frac{H(z)/(1+z)}{\mathrm{d} v_\|/\mathrm{d} r_\|}\right)\,,
\end{aligned}
\end{equation}
where $\delta_b$ is the baryon overdensity field, $x_{\rm HI}$ is the fraction of hydrogen that is neutral, $T_{\rm CMB}(\nu)$ is the CMB temperature at frequency $\nu$, $\mathrm{d} v_\|/\mathrm{d} r_\|$ is the gradient of the proper velocity along the line-of-sight direction, and $T_{\rm spin}$ is defined as the ratio of the occupancy of the spin-1 and spin-0 ground states of the neutral hydrogen:
\begin{align}
    \frac{n_1}{n_0} = 3\exp\left(-T_* / T_{\rm spin}\right)\text{  with $T_*=0.0681$ K}\,.
\end{align}
For the redshift range of interest for this work, it is safe to assume that $T_\mathrm{spin} \gg T_\mathrm{CMB}$ and ignore the term $(1-T_\mathrm{CMB}/T_\mathrm{spin})$. The remaining quantities in Eq.\,\eqref{eq:21cm_brightness} are obtained using gas properties sampled on a $512^3$ regular Cartesian grid \citep{Garaldi2023:THESAN_Data_Release}\footnote{\href{https://www.thesan-project.com/thesan/cartesian.html}{https://www.thesan-project.com/thesan/cartesian.html}}. These quantities are binned in redshift space to take into account redshift-space distortions. We note that \textsc{thesan} was run under the cosmological parameters from \citet{Planck15}. In particular, we have $h = 0.6774$, $\Omega_m = 0.3089$, $\Omega_\Lambda = 0.6911$, $\Omega_b = 0.0486$, $\sigma_8 = 0.8159$, and $n_s =0.9667$. All cosmological calculations in this work assume the same cosmology.

\begin{figure*}[tbh]
    \centering
    \epsscale{1.175}
    \plotone{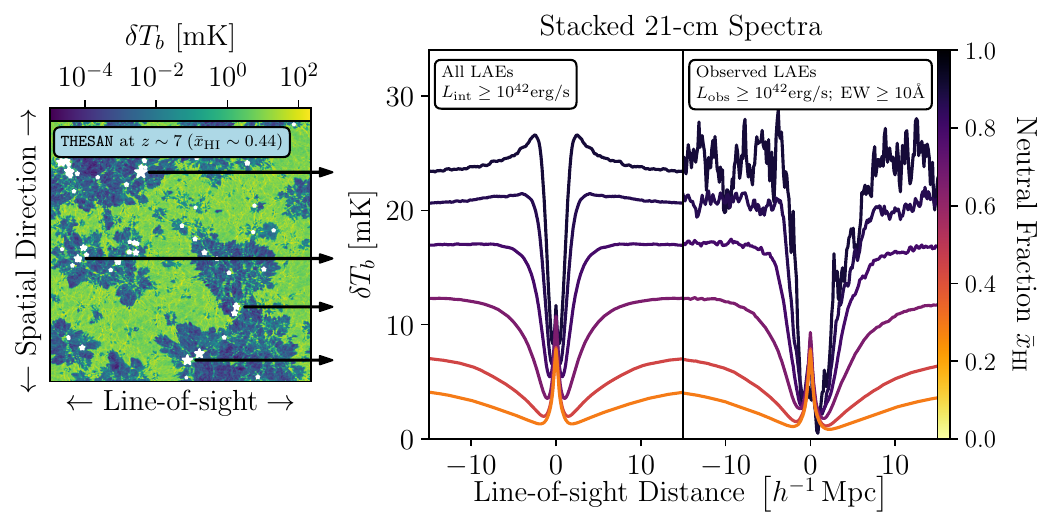}
    \caption{\textit{Left}: Snapshot of 21-cm brightness temperature from \textsc{thesan} at redshift 7 with LAEs marked in white stars. We draw 21-cm spectra along the line of sight from each LAE to form a template for the stacked 21-cm signal. \textit{Right}: Stacked 21-cm spectra around LAEs that are intrinsically bright (left) and can be observed by a fiducial ground-based spectroscopy survey (right). Different curves mimic a different reionization history which predicts a different global neutral fraction at redshift $z\sim 7$. \label{fig:signal_model} }
\end{figure*}

\subsection{LAE Properties} \label{subsec:LAE_properties}

Because the neutral IGM is optically thick at the Ly$\alpha$ wavelength, it is crucial to take into account the correlation between observed LAEs and their surrounding IGM. Here, we describe how we obtain the observed Ly$\alpha$ properties through different lines of sight for each LAE. These observed properties are empirically calibrated to ensure that the resulting Ly$\alpha$ luminosity functions match the observations. Details of this process will be described in the upcoming work \citet{Neyer2025}.

First, intrinsic properties of the Ly$\alpha$ emission from galaxies are calculated directly from \textsc{thesan}. The intrinsic Ly$\alpha$ luminosity $L_{\alpha,\rm int}$ incorporates contributions from recombinations, collisional excitations, and unresolved HII regions \citep{Smith20222:THESAN_Lya}. The frequency-dependent Ly$\alpha$ transmission as the photons pass through the local IGM is also accurately captured through an effective absorption treatment with continuous Doppler shifting, i.e. $\mathcal{T}_{\rm IGM}(\nu) = \exp[-\tau(\nu)]$, extracting sightlines with the \textsc{colt} code \citep{Smith2015:COLT, Smith2019:COLT, Smith2022:COLT}. To account for unresolved galaxy-scale phenomena including dust, outflows, and other effects from the interstellar and circumgalactic medium, an idealized model for a Ly$\alpha$ point source surrounded by an expanding or contracting gas cloud is applied and calibrated to observational constraints on the Ly$\alpha$ luminosity functions at $z=5.7$ and $z=6.6$ \citep{Ouchi2008:Subaru_LF, Ouchi2010:Subaru_LF, Konno2018:Subaru_LF}. The resulting fit is used to calculate the escape fraction $f_{\rm esc}$ and calibrate the spectral profile for each galaxy. Together with the sightline-dependent IGM transmission $\mathcal{T}_{\rm IGM}$, these quantities are combined to derive observed Ly$\alpha$ luminosities $L_{\alpha,\rm obs}$ and equivalent widths (EW) for each sightline from a galaxy. The observed luminosity is calculated as $L_{\alpha,\rm obs} = f_{\rm esc} \times \mathcal{T}_{\rm IGM} \times L_{\alpha,\rm int}$ and the equivalent width is calculated as ${\rm EW} = L_{\alpha,\rm obs} / L_{\lambda,\mathrm{cont}}$, where $L_{\lambda,\mathrm{cont}}$ is the specific luminosity of the stellar continuum surrounding the Ly$\alpha$ emission line.


\subsection{Signal Template} \label{subsec:signal_template}

To derive a signal template for the stacked 21-cm spectrum around LAEs, we utilize 768 lines of sight (in Healpix directions) from each galaxy in \textsc{thesan}. Based on the observed LAE properties along each line of sight, we can implement the selection of any galaxy survey and derive the signal template by stacking the 21-cm spectra only when we can observe an LAE. 

As an example, we consider we have a sample of LAEs at $z_{\mathrm{LAE}}\sim7$. The leftmost panel of \autoref{fig:signal_model} shows the 21-cm brightness temperature in \textsc{thesan} at this redshift with LAEs marked in white stars. What does the stacked 21-cm spectrum around these LAEs look like? The right two panels of \autoref{fig:signal_model} show how the result differs if we select LAEs based on their intrinsic versus observed properties. Here, a positive value in the $x$-axis indicates the direction toward the observer. 

We note that the prescriptions in \textsc{thesan} give rise to a particular model of reionization history. To generalize our signal template to account for different reionization scenarios, we calculate the stacked 21-cm spectra with LAEs at various snapshots with different $\bar{x}_\mathrm{HI}$. Following Eq.\,\eqref{eq:21cm_brightness}, we scale the resulting 21-cm spectra by $\sqrt{(1+z_{\mathrm{LAE}})/(1+z_{\mathrm{snap}})}$ where $z_{\mathrm{snap}}$ is the redshift of each snapshot. Hence, each curve in the right two panels of \autoref{fig:signal_model} corresponds to a stacked 21-cm spectrum at $z\sim7$ assuming a different global neutral fraction. Here, we also show the difference between stacking 21-cm spectra around intrinsically bright LAEs (left) versus observed LAEs (right). In both cases, the brightness temperature dips around the center as the IGM are mostly ionized there, except for a small emission peak from the neutral hydrogen within the galaxies. However, if we stack around intrinsically bright LAEs, the absorption troughs do not go all the way to zero, especially when the IGM is more opaque (higher $\bar{x}_\mathrm{HI}$). This is because not every LAE resides in an ionized bubble, whereas the observed LAEs are guaranteed to be surrounded by a more transparent IGM. A major feature of this result is that the amplitude of the stacked 21-cm spectrum becomes a direct tracer of the global neutral fraction $\bar{x}_\mathrm{HI}$. This coincides with the finding in \citet{Hutter2023:21cm_LAE} as the amplitude of the stacked spectrum is approximately equivalent to the two-point correlation function between 21-cm and galaxies at very small scales. Moreover, we see that the absorption profiles are largely symmetric in the left panel. The asymmetry in the rightmost panel arises because observed LAEs preferentially reside in the back side of ionized regions. Observationally speaking, while stacking around a sample of LAEs yields spectra like those on the right-hand side, we can obtain signal that bear more resemblances to the template on the left-hand side if we stack around galaxies detected through other emission lines such as [OIII].

Here, we choose a Ly$\alpha$ luminosity threshold of $10^{42}\,\mathrm{erg}/\mathrm{s}$ as an example. At $z\sim7$, this roughly corresponds to a survey with a flux limit of $10^{-18}\,\mathrm{erg}/\mathrm{s}/\mathrm{cm}^2$. This can be achieved by large ground-based spectroscopy \citep[e.g.,][]{Hu2017:LAGER_specz, Yang2019:LAGER_specz, Harish2022:LAGER_specz} and is approximately an order of magnitude deeper than what the Roman grism survey will achieve \citep{ROTAC_Report}. Changing the selection criteria in either Ly$\alpha$ luminosity or the equivalent width modifies the morphology of our signal, but the variation is not significant given the sensitivity of current 21-cm experiments. 

\section{Forecast}\label{sec:results}
\subsection{Setup} \label{subsec:forecast_setup}
\begin{figure}[t]
    \centering
    \epsscale{1.175}
    \plotone{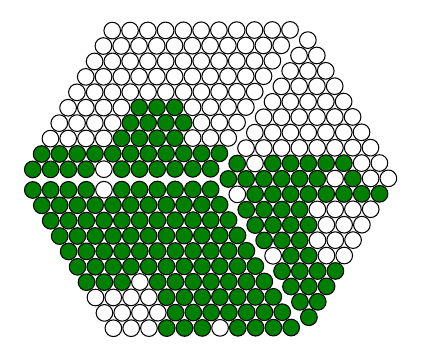}
    \caption{Layout of the 320 core antennas of HERA used in this forecast. The 172 antennas marked in green have been taking data since 2022 and are used as a baseline configuration to investigate the prospect of cross-correlation with existing HERA data. \label{fig:hera_layout} }
\end{figure}

In this work, we focus on forecasting the detectability of a cross-correlation signal with HERA. HERA is a 350-element radio interferometer located in the Karoo desert in South Africa. In particular, we consider only the 320 elements that form a compact core array with dishes that almost touch each other. The maximum inter-antenna distance is 292 meter and the shortest baseline is 14.6 meter. The antenna configuration can be seen in \autoref{fig:hera_layout}.

Currently, science data are being taken by a subset of commissioned antennas, while new antennas are continuously being added. In particular, 172 antennas marked in green in \autoref{fig:hera_layout} have been taking data since 2022. We use this subset of antennas to form a conservative forecast to investigate whether a cross-correlation detection is possible with data that have already been taken by HERA to date. 

To simulate the signal observed in real HERA images, we take the stacked signal template derived in Sec.\,\ref{subsec:signal_template} and generate simulated visibility following Eq.\,\eqref{eq:measurement_eq}. In this work, we assume that each antenna has an Airy beam profile,
\begin{equation}
    B\left(\hat{\mathbf{s}}(\theta, \phi); \nu\right) = \left[\frac{2J_1(2\pi\nu a \sin\theta/c)}{2\pi\nu a \sin\theta/c}\right]^2\, ,
\end{equation}
where $J_1$ is the Bessel function of the first kind, $\theta$ is the zenith angle, and $a$ is the aperture radius which we set to be six meters to mimic an underillumination HERA dish \citep{Neben2016:HERA_Beam, Orosz2019:BeamVariation, HERA2017:PhaseI_Overview}. The visibility from each baseline is then filtered according to the procedure outlined in Sec.\,\ref{subsec:foreground_filter} and \autoref{appendix:fg_filter}. The foreground-filtered visibilities are then combined and map to the image space following Eq.\,\eqref{eq:mapping_from_vis} in which the normalization is chosen to be as Eq.\,\eqref{eq:normalization}. One important feature is that this entire procedure---from signal to mock image---is linear. Hence, mapping a stacked signal is equivalent to mapping individual galaxies and stacking them afterward. We denote the observed stacked signal as $s_\mathrm{obs}(\nu)$. 

Another source of uncertainty in the signal comes from the redshift of the LAEs. In order to perfectly align and stack the 21-cm spectra, we need to know precisely the redshift of these LAEs, at least to the precision that matches the frequency resolution of the radio instrument. At redshift 7, the frequency resolution of HERA corresponds to a redshift uncertainty $\sigma_z\approx0.005$. This matches well with the uncertainties provided by a space-based grism or a large ground-based spectroscopy. Here, we consider three different redshift uncertainties, $\sigma_z = 0.001, 0.01, 0.1$. These redshift uncertainties are incorporated by perturbing the location of the galaxies when deriving the stacked signal template in Sec.\,\ref{subsec:signal_template}.

\begin{figure}[tb]
    \centering
    \epsscale{1.175}
    \plotone{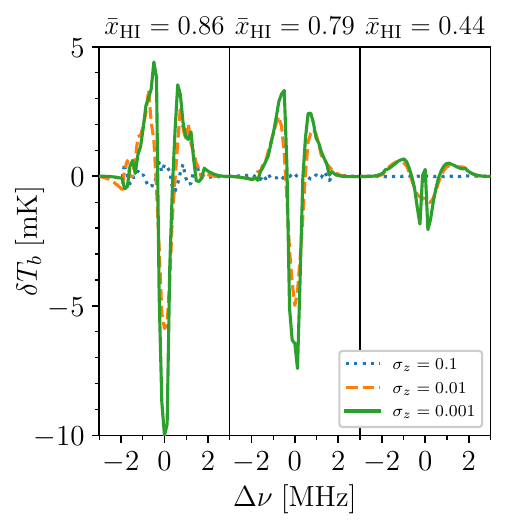}
    \caption{Stacked 21-cm brightness temperature signal as observed by HERA assuming different neutral fraction $\bar{x}_\mathrm{HI}$ at $z\sim7$. Different lines correspond to different redshift uncertainties for the LAEs. These signals have gone through the foreground filtering procedure as described in Sec.\,\ref{subsec:foreground_filter} and \autoref{appendix:fg_filter}. \label{fig:signal_filtered_obs} }
\end{figure}

\autoref{fig:signal_filtered_obs} shows the resulting stacked 21-cm signal as observed by HERA. These signals have been foreground-filtered; hence, the smooth component of the signal is removed. The three panels indicate the different signal strength if the global neutral fraction $\bar{x}_\mathrm{HI}$ is $0.86$, $0.79$, or $0.44$ at redshift $\sim7$. The default reionization history in \textsc{thesan} predicts $\bar{x}_\mathrm{HI} = 0.44$ at $z\sim7$. A higher neutral fraction means bigger contrast between the average brightness temperature of the IGM and the ionized bubble (which has a brightness temperature around $0$). In each panel, the three different curves show the effect of redshift uncertainties on the stacked signal. The cross-correlation signal is maximized with minimal redshift uncertainties. We see that the signal almost vanishes for $\sigma_z = 0.1$ (dotted blue), which is typical for a photometric redshift estimate. Spectroscopy confirmation of these LAEs is therefore necessary for a successful cross-correlation detection through stacking along the line-of-sight direction. 

Throughout this work, we focus on forecasting the cross-correlation signal around redshift $7$. This is the redshift where most LAEs are currently being identified on the ground due to sky lines. Searching for such a signal at higher redshift might be easier due to the increased signal strength. At the same time, a more opaque IGM makes it harder to identify a large sample of LAEs. With the launch of the \textit{James Web Space Telescope}, we are starting to more evenly sample LAEs at even higher redshifts \citep[e.g.,][]{Kumari2024:JADES_LAE, Tang2024:JWST_LAEs, Witstok2025:z_13_LAE}. With upcoming wide-area grism surveys on \textit{Euclid} \citep{Euclid} and \textit{Roman} \citep{Roman}, it will soon be possible to probe cross-correlation signals across the entire reionization history.

We note that while this forecast focuses on HERA, the result presented here can be generalized to other experiments. As we do not attempt to map structures in the spatial direction, the exact layout of antennas is less relevant to our sensitivity forecast. This is seen in Eq.\,\eqref{eq:pixel_variance} as the noise level in a given pixel in the 21-cm image depends only on the total number of baselines in the experiment. For experiments with steerable antennas, the dilution of signal from antenna's primary beam in Eq.\,\eqref{eq:pt_source_SN} can be avoided by tracking a source. The forecast for a transit telescope therefore forms a conservative lower bound for an equivalent tracking experiment. For context, the full HERA core array has 51360 baselines, while the 172 antennas that are currently taking data form 14706 baselines.

\begin{figure*}[tbh]
    \centering
    \epsscale{1.175}
    \plotone{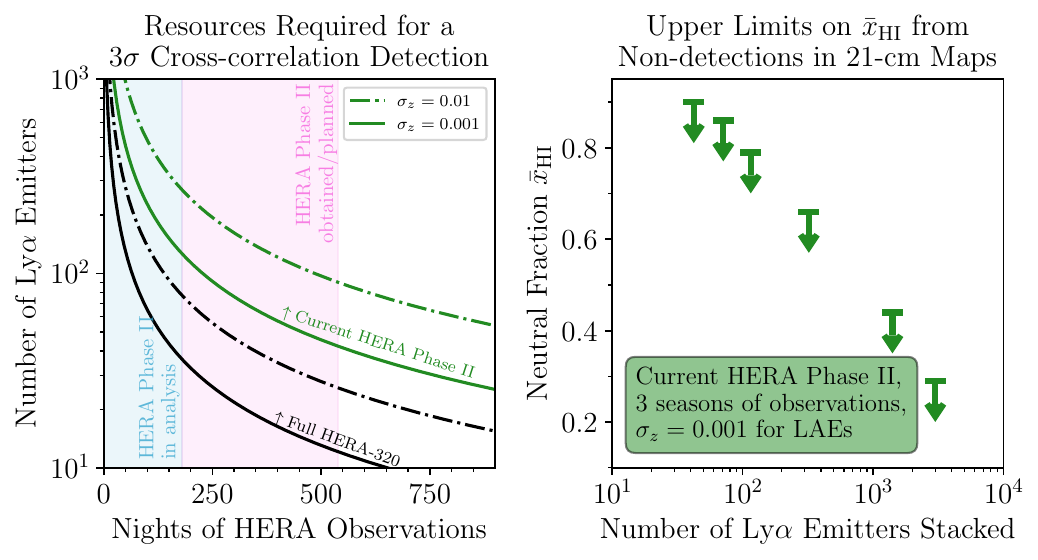}
    \caption{$\textit{Left}$: Minimum resources required for a $3
    \sigma$ cross-correlation detection with HERA. The green lines are for the current HERA layout while the black lines are for the full HERA-320 (see \autoref{fig:hera_layout}). $\textit{Right}$: Upper limits on the global neutral fraction $\bar{x}_\mathrm{HI}$ at $z\sim 7$ derived from non-detections of any cross-correlation signal as a function of number of LAEs stacked. This assumes the 21-cm maps are generated from the three seasons ($\sim$540 nights) of HERA observations that have already been taken, and the data are dominated by thermal noise instead of systematic effects. \label{fig:forecast} }
\end{figure*}

\subsection{Results} \label{subsec:results}

The main question we focus on in this work is: what are the observing resources required to make a significant cross-correlation detection through stacking 21-cm spectra around LAEs? Using the observed signal template $s_\mathrm{obs}(\nu)$ derived in Sec.\,\ref{subsec:forecast_setup}, we calculate the signal-to-noise ratio for the cross-correlation to be 
\begin{equation}
    \mathrm{SNR}\coloneqq\sqrt{\sum_{ij} s_\mathrm{obs}(\nu_i) C^{-1}_{ij} s_\mathrm{obs}(\nu_j)}\, ,
\end{equation}
where the covariance matrix $\vect{C}$ is given in Eq.\,\eqref{eq:img_covariance}. The covariance matrix takes into account the correlation between different frequency channels introduced by foreground filtering, and its variance is determined by the amount of observing resources used to reduce the noise level. 

The noise level in the stacked 21-cm images is determined by three factors: the integration time of each nightly observation, the number of nights we can observe each object, and the number of objects (LAEs) available for stacking. Based on the results described in Sec.\,\ref{subsec:statistics}, the maximum sensitivity we can achieve around each galaxy occurs when it is observed for roughly 45 minutes as it transits the zenith. This gives an effective coherent averaging time of 25 minutes. Ideally, each source can be observed repeatedly every night for half the year. We present our forecast in terms of the number of nights and LAEs required to obtain a significant cross-correlation signal. 

The top panel of \autoref{fig:forecast} shows the minimum amount of resources required for a $3\sigma$ cross-correlation detection with HERA. This assumes the signal to be the strongest as seen in \autoref{fig:signal_filtered_obs}. The green lines are for the current HERA layout, while the black lines are for the full HERA-320 (see \autoref{fig:hera_layout}). Lower redshift uncertainties also make the signal stronger and easier to be detected. Since 2022, three seasons of HERA Phase II data have been taken. With roughly $540$ nights of available data, a cross-correlation detection starts to become possible with around $43$ ($90$) LAEs assuming $\sigma_z = 0.001~(0.01)$. 

The fiducial reionization model adopted in \textsc{thesan} predicts $\bar{x}_\mathrm{HI}\sim0.44$ at $z\sim7$. In this case, around $1000$ LAEs are required to detect the cross-correlation signal with the current HERA dataset. However, we note that the neutral fraction changes rapidly as we move to higher redshift while the properties of the 21-cm maps remain roughly the same. Moving to $z\sim8$, we need about $300$ LAEs to make a detection with the same dataset. In reality, as we have access to 21-cm information across a wide range of redshift, we can make inference at any redshift bin where we have a significant sample of LAEs.  

Even in the event where there are not enough LAEs to make a detection, we can turn a non-detection with a given number of LAEs into an upper limit on the global neutral fraction. This is because the signal strength is proportional to the global neutral fraction $\bar{x}_\mathrm{HI}$. A non-detection at a given sensitivity suggests that the neutral fraction must be lower than a certain level. The bottom panel of \autoref{fig:forecast} shows an example of upper limits that can be derived from existing HERA data, if we do not measure a cross-correlation signal after stacking a given amount of spectroscopically confirmed LAEs. Such a result will be a unique constraint on the reionization history measured directly from the IGM. 

\section{Conclusion}\label{sec:conclusion}

We have presented a detailed analysis of a direct way to detect a cross-correlation signal between galaxies and the 21-cm field---through stacking 21-cm image cubes around Lyman-alpha emitters (LAEs). Detections of the cross-correlation signal are of increasing importance as 21-cm experiments approach the sensitivity to detect an auto power spectrum. Cross-correlations can be crucial to verify any 21-cm auto power spectra detection is cosmological in nature. Moreover, we have shown that even an upper limit on the cross-correlation signal provides additional cosmological information. Current limits on the 21-cm auto power spectra are only sensitive to the heating of the intergalactic medium (IGM). By combining with information provided by high-redshift galaxies, a simple cross-correlation through stacking can already provide additional constraints on the reionization history. 

One of the foremost requirements to make any inference in 21-cm cosmology is to understand the statistical properties of the data. In the case of stacking, the 21-cm image cubes. In this work, we have derived the statistical properties of foreground-filtered 21-cm maps generated with a series of linear operations. We choose to work with a linear map-making and foreground-filtering algorithm to ensure that we can correctly estimate and propagate the statistical properties of our maps. 

On the theory side, we have derived a signal template using state-of-the-art radiation-magneto-hydrodynamic cosmological simulations \textsc{thesan}. This provides a physically driven model that connects galaxies to their surrounding ionized bubbles. Our results also take into account the selection effect of LAEs. The correlation between the observed Lyman-alpha properties of a galaxy and its surrounding IGM is accounted for with radiative transfer modeling. Unresolved galaxy-scale phenomena are further calibrated to observational constraints on the Lyman-alpha luminosity functions at high redshift. An important feature we have confirmed is that the stacking signal is proportional to the averaged neutral fraction of the universe.

In conclusion, our forecast suggests that we are in a position to place significant constraints on the reionization history with existing HERA data through stacking. A sample of around 50 (100) LAEs with redshift uncertainties of 0.001 (0.01) is sufficient to begin with. Around $300$ ($1000$) LAEs are required to make a detection with existing HERA dataset if the neutral fraction is around $0.6$ ($0.4$) at $z\sim7$. Such a sample of LAEs could soon be available with upcoming space-based grism surveys. The prospect of a detection will improve in the meantime with more commissioned antennas and continuous observations. Significant improvements in detectability can also be achieved with advanced analysis techniques in inferring cosmological modes within the foreground wedge \citep[e.g.][]{Chen2025:21cmEFT, Qin2025:21cmEFT}, or through designing experiments that reduce the foreground wedge \citep[e.g.][]{MacKay:RULES}. These methods can alleviate the amount of signal loss during foreground mitigation. 

Lastly, we discuss some of the limitations of this work. On the theory side, our signal template is slightly limited by the box size of the simulation. At higher neutral fractions, the number of LAEs that pass the selection criteria is low, leading to a nosier signal template, as can be seen in \autoref{fig:signal_model}. With larger boxes of radiation-hydrodynamic simulations soon to be available, we expect a significant improvement in comic variance uncertainties in our signal template. On the observation side, an important assumption we have made is that the 21-cm image cubes are free of systematic effects after foreground filtering. This might not be true in the real world due to imperfect or insufficient characterizations of the instrument, especially after significantly averaging down the data. Another source of systematic uncertainty that we omit is the contamination from surrounding emissions. The brightness temperature at a given pixel in our map is a convolution of all pixels on the sky with the point spread function (synthesized beam) of the array. If the point spread function varies smoothly as a function of frequency, this should only add smooth contamination to our signal and would be removed by foreground filtering. However, careful consideration of these systematic effects is necessary when handling real data. 

\begin{acknowledgments}
We thank James Rhoads, Sangeeta Malhorta, Isak Wold, Crist\'obal Andr\'es Moya Sierralta, Adrian Liu, Julian Mu\~noz, Dominika  {\v{D}}urov{\v{c}}{\'\i}kov{\'a} for helpful discussions and comments on this paper. This research is funded in part by the Gordon and Betty Moore Foundation through Grant GBMF5212 to the Massachusetts Institute of Technology. AS acknowledges support through HST AR-17859, HST AR-17559, and JWST AR-08709. MV acknowledges support through NSF AAG AST-2408412, JWST JWST-AR-04814, NASA ATP 21-ATP21-0013, NASA ATP 21-ATP21-0013 and NSF AAG AST-2307699.
\end{acknowledgments}

\software{
\texttt{colossus} \citep{Colossus}, 
\texttt{pyuvdata} \citep{Hazelton2017:pyuvdata, Keating2025}, 
\texttt{direct\_optimal\_mapping} (\url{https://github.com/HERA-Team/direct_optimal_mapping},
\texttt{astropy} \citep{astropy},  
\texttt{numpy} \citep{numpy}, 
\texttt{scipy} \citep{scipy}, 
\texttt{jupyter} \citep{Kluyver2016jupyter},
\texttt{matplotlib} \citep{matplotlib}, 
\texttt{seaborn} \citep{seaborn}.
}


\appendix
\section{Foreground Filtering with Discrete Prolate Spheroidal Sequence}
\label{appendix:fg_filter}

In this appendix, we describe the detailed procedure for generating the linear operator $\vect{\mathcal{O}^\mathrm{fil}}$ that performs foreground filtering using the discrete prolate spheroidal sequence (DPSS). DPSS is the set of eigenvectors to the prolate matrix $\vect{B}$ where
\begin{equation}
\label{eq:prolate_matrix}
    B_{ij} = \frac{\sin{2\pi T(\nu_i - \nu_j)}}{\pi (\nu_i - \nu_j)}\, .
\end{equation}
Here, $T$ denotes the baseline-dependent delay range below which we want to filter out the smooth foreground component. In this work, for a baseline vector $\vect{b}$, we choose $T=|\vect{b}|/c$ to remove all modes within the foreground wedge. We note that in practice, one might choose a slightly larger $T$ to also filter out foreground contamination into the EoR window due to various systematic effects. 

To fit and filter the smooth foreground, we use all eigenvectors $f_i(\nu)$ of Eq.\,\eqref{eq:prolate_matrix} with eigenvalues $\lambda_i \geq 10^{-12}$. The eigenvalues of the prolate matrix are always between $0$ and $1$ and denote how localize each eigenvector is within the given delay range $[-T, +T]$ ($1$ being completely localized). A lower eigenvalue cut here ensures that we have a more complete basis, which reduces the amount of residual foreground. Although this also gives us some eigenvectors that could filter out signal in the high delay EoR window, the number of these eigenvectors is fairly limited \citep{Karnik2020:DPSS_Bounds} and the signal loss is carefully quantified in our work.

Once a set of basis $\{f_i\}_{i=1}^N$ is chosen, we fit the smooth foreground by solving the linear system
\begin{equation}
    \vect{v}_\mathrm{obs} = \vect{A}\vect{\alpha} + \vect{n}\,,
\end{equation}
where $\mathbf{v}_\mathrm{obs}$ is the observed visibility of a given baseline, $A_{ij} = f_j(\nu_i)$ is the \textit{design matrix}, $\vect{\alpha}$ is the DPSS coefficient we wish to solve, and $\mathbf{n}$ is the instrumental thermal noise that can be modeled with Eq.\,\eqref{eq:noise_matrix}. The maximum likelihood estimator of $\vect{\alpha}$ is then
\begin{equation}\label{eq:b_ML}
    {\vect{\hat{\alpha}}} = (\mathbf{A}^\dagger\mathbf{N}^{-1}\mathbf{A})^{+} \mathbf{A}^\dagger\mathbf{N}^{-1}\mathbf{v}_\mathrm{obs}\,,
\end{equation}
where $\vect{M}^+$ denotes the Moore–Penrose pseudo-inverse of a matrix $\vect{M}$. By subtracting the best-fit smooth foreground, we obtain the filtered visibility $\vect{v}^\mathrm{fil}$ 
\begin{equation}
    \vect{v}^\mathrm{fil} \coloneqq \vect{v}_\mathrm{obs} - \vect{A}\vect{\hat{\alpha}} = \left[\vect{I} - \vect{A}(\mathbf{A}^\dagger\mathbf{N}^{-1}\mathbf{A})^{+} \mathbf{A}^\dagger\mathbf{N}^{-1}\right]\vect{v}_\mathrm{obs}\,.
\end{equation}
Hence, $\left[\vect{I} - \vect{A}(\mathbf{A}^\dagger\mathbf{N}^{-1}\mathbf{A})^{+} \mathbf{A}^\dagger\mathbf{N}^{-1}\right]$ is the linear foreground filter operator in Eq.\,\eqref{eq:fg_filt}. In this work, we filter the visibility across a wide frequency range to reduce signal attenuation \citep{Ewall-Wice2021:DAYNENU, Kern2021:GPR}. For the results presented in this work, foreground is fit with data between $108.08$ to $234.30$ MHz which correspond to the full range of HERA data above the FM radio band. 


\bibliography{ref}{}
\bibliographystyle{aasjournalv7}



\end{document}